# A new type of non-Hermitian phase transition in open systems far from thermal equilibrium

T. T. Sergeev[1,2,3], A. A. Zyablovsky[1,2,3,4], E. S. Andrianov[1,2,3], A. A. Pukhov,[2,3] Yu. E. Lozovik[1,2,5], A. P. Vinogradov[1,2,3]

[1]Dukhov Research Institute of Automatics (VNIIA), 127055, 22 Sushchevskaya, Moscow, Russia

[2]Moscow Institute of Physics and Technology, 141700, 9 Institutskiy per., Moscow, Russia

[3]Institute for Theoretical and Applied Electromagnetics, 125412, 13 Izhorskaya, Moscow, Russia

[4]Kotelnikov Institute of Radioengineering and Electronics RAS, 11-7 Mokhovaya, Moscow 125009, Russia

[5]Institute of Spectroscopy Russian Academy of Sciences, 108840, 5 Fizicheskaya, Troitsk, Moscow, Russia

We demonstrate a new type of non-Hermitian phase transition in open systems far from thermal equilibrium, which takes place in coupled systems interacting with reservoirs at different temperatures. The frequency of the maximum in the spectrum of energy flow through the system plays the role of the order parameter, and is determined by an analog of the $\varphi^4$-potential. The phase transition is exhibited in the frequency splitting of the spectrum at a critical point, the value of which is determined by the relaxation rates and the coupling strengths. Near the critical point, fluctuations of the order parameter diverge according to a power law. We show that the critical exponent depends only on the ratio of reservoir temperatures. This dependence indicates the non-equilibrium nature of the phase transition at the critical point. This new non-Hermitian phase transition can take place in systems without exceptional points.

**Introduction**

Non-Hermitian systems possess many unusual properties [1,2], and one of the most interesting features of these systems is the presence of exceptional points (EPs) [3,4]. An EP is a spectral singularity in the parameter space of a non-Hermitian system in which two or more eigenstates become linearly dependent and their eigenfrequencies coalesce [2-4]. When an EP is crossed, the eigenstates of the non-Hermitian system change; for example, the symmetry of the eigenstates may be spontaneously broken [4-7]. Of the non-Hermitian systems that have EPs, PT-symmetric systems are notable [3-7]. In these systems, the EP separates areas of parameter space with PT-symmetric and non-PT-symmetric eigenstates [4-7]. Other examples of non-Hermitian systems with an EP include strongly coupled cavity-atom systems [3,8], polariton systems [9-11], optomechanical systems [12,13], and even conventional laser systems [14]. Non-Hermitian systems with EPs have a number of unique properties, due to which they have found many applications [3,4,6,7]. For example, they are used to enhance the sensitivity of laser gyroscopes [15] and sensors [16-19], to achieve single-mode lasing in multimode systems [20,21] and directional lasing [22], and to control the lasing threshold [23-25]. In addition, a laser system with an EP can be used to achieve lasing with a negative population inversion [26,27].

Due to the changes in the eigenstates of a non-Hermitian system at an EP, the transition through the EP is often associated with the point of non-Hermitian phase transition [3,4]. In experiments, non-Hermitian phase transitions are observed as frequency splitting in the system spectrum [10,28-30]. However, relaxation and noise tend to shift this splitting point in the spectrum away from the EP [31-34]. An analogy with a phase transition is further complicated by the fact that non-Hermitian systems are generally non-equilibrium systems. Indeed, non-Hermitian systems with EPs inevitably consist of several subsystems that interact with the different reservoirs. These reservoirs of various types (e.g., photon and phonon reservoirs) are not necessarily in thermal equilibrium with each other, which moves the system far from thermal equilibrium and results in an energy flow between the reservoirs.

In this letter, we use the example of two coupled oscillators that interact with reservoirs at different temperatures to demonstrate a new type of non-Hermitian phase transition. This transition leads to frequency splitting in the spectrum of energy flow between the reservoirs. We show that this spectrum is determined by an analog of the $\varphi^4$-potential, and that it exhibits frequency splitting at a critical point (CP) above which there are two minima in the potential. Using an analogy with the theory of second-order phase transitions, we determine the frequency of the maximum in the spectrum as an order parameter for this non-Hermitian transition. We demonstrate that near the CP, fluctuations of the order parameter diverge according to a power law. The corresponding critical exponent depends only on the ratio of the temperatures of the reservoirs, and remains unchanged when the temperatures of all the reservoirs change by the same factor. This indicates the non-equilibrium nature of the phase transition at the CP. It is remarkable that the CP does not depend on the reservoir temperatures. Like an EP, the CP depends only on the relaxation rates of the system and the coupling strengths between its subsystems.

Our results open the way for observing non-Hermitian phase transitions in systems far from thermal equilibrium. The conditions for a CP in this transition differ from the conditions for an EP, thus making it possible to observe non-Hermitian phase transitions in systems without an EP, e.g., in a system of two coupled oscillators with the same relaxation rate but with different reservoir temperatures.

**2. The model**

We consider two coupled oscillators interacting with their own reservoirs. The reservoirs are in thermodynamic equilibrium with temperatures $T_1$ and $T_2$, respectively. By eliminating the degrees of freedom of the reservoir using the Born-Markovian approximation [35,36], we can obtain the equations for the oscillator amplitudes [35,36]:

$$\frac{d}{dt}\begin{pmatrix} a_1 \\ a_2 \end{pmatrix} = \begin{pmatrix} -\gamma_1 & -i\Omega \\ -i\Omega & -\gamma_2 \end{pmatrix}\begin{pmatrix} a_1 \\ a_2 \end{pmatrix} + \begin{pmatrix} \xi_1 \\ \xi_2 \end{pmatrix}. \quad (1)$$

Here, $\vec{a} = (a_1, a_2)^T$ is a vector of the amplitudes of the first and second oscillators, respectively; $\Omega$ is the coupling strength between the oscillators; $\gamma_{1,2}$ are the relaxation rates of the oscillators; and $\vec{\xi} = (\xi_1 \ \xi_2)^T$ is a noise term that always appears in the Born-Markovian approximation together with the relaxation terms [35,36]. According to the fluctuation-dissipation theorem [35,37], the noise terms obey the following correlation properties

$$\langle \vec{\xi} \rangle = 0, \quad \langle \vec{\xi}^*(t+\tau) \vec{\xi}^T(t) \rangle = 2\hat{D}\delta(\tau), \tag{2}$$

where $\hat{D} = \begin{pmatrix} \gamma_1 T_1 & 0 \\ 0 & \gamma_2 T_2 \end{pmatrix}$ is a diffusion matrix. For brevity, we denote the matrix on the right-hand side of Eq. (1) as $\hat{M} = \begin{pmatrix} -\gamma_1 & -i\Omega \\ -i\Omega & -\gamma_2 \end{pmatrix}$.

We calculate the spectrum of stationary fluctuations of the system using the Wiener-Khinchin theorem [38,39]:

$$\hat{S}(\omega) = \frac{1}{2\pi} \int_{-\infty}^{+\infty} d\tau \exp(-i\omega\tau) \langle \vec{a}^*(\tau) \vec{a}^T \rangle_{st}, \tag{3}$$

where $\langle \vec{a}^*(\tau) \vec{a}^T \rangle_{st}$ is a matrix of two time correlators that is determined using regression theorem as follows (see Supplementary Materials for details):

$$\langle \vec{a}^*(\tau) \vec{a}^T \rangle_{st} = \begin{cases} \exp(\hat{M}^* \tau) \langle \vec{a}^* \vec{a}^T \rangle_{st}, & \tau > 0 \\ \exp(-\hat{M}^T \tau) \langle \vec{a}^* \vec{a}^T \rangle_{st}, & \tau < 0 \end{cases}.$$

The averages $\langle \vec{a}^* \vec{a}^T \rangle_{st} = \langle \vec{a}^*(\tau=0) \vec{a}^T \rangle_{st}$ obey the following condition [35] (see also the Supplementary Materials):

$$\hat{M}^* \langle \vec{a}^* \vec{a}^T \rangle_{st} + \langle \vec{a}^* \vec{a}^T \rangle_{st} \hat{M}^T + 2\hat{D} = 0. \tag{4}$$

It should be noted that the noise results in nonzero stationary values for the oscillator energies $\langle a_1^* a_1 \rangle_{st}$ and $\langle a_2^* a_2 \rangle_{st}$ as well as the value $\langle a_1^* a_2 \rangle_{st}$. The real part, $\text{Re}\langle a_1^* a_2 \rangle_{st}$, represents the interaction energy between the oscillators, while the imaginary part, $\text{Im}\langle a_1^* a_2 \rangle_{st}$, represents the energy flow from the first oscillator to the second.

Using the Eqs. (3) and (4) we obtain the following expression for the spectrum [Carmichael]:

$$\hat{S}(\omega) = \frac{1}{\pi} (\hat{M}^* - i\omega \hat{I})^{-1} \hat{D} (\hat{M}^T + i\omega \hat{I})^{-1}. \tag{5}$$

Using the definitions of matrices $\hat{M}$ and $\hat{D}$, we obtain:

$$\hat{S}(\omega) = \frac{1/\pi}{\Phi(\omega)} \begin{pmatrix} \gamma_1 T_1 (\gamma_2^2 + \omega^2) + \gamma_2 T_2 \Omega^2 & i\Omega(\gamma_2 T_2(\gamma_1 + i\omega) - \gamma_1 T_1(\gamma_2 - i\omega)) \\ i\Omega(\gamma_1 T_1(\gamma_2 + i\omega) - \gamma_2 T_2(\gamma_1 - i\omega)) & \gamma_2 T_2 (\gamma_1^2 + \omega^2) + \gamma_1 T_1 \Omega^2 \end{pmatrix}, \tag{6}$$

where $\Phi(\omega) = (\gamma_1\gamma_2 + \Omega^2)^2 - 2(\Omega^2 - \Omega_{cr}^2)\omega^2 + \omega^4$ and $\Omega_{cr}^2 = (\gamma_1^2 + \gamma_2^2)/2$. The diagonal elements of $\hat{S}(\omega)$ determine the spectra of the fluctuations of the first and second oscillators. The non-diagonal elements determine the spectrum of interaction between the oscillators. The real part represents the spectrum of fluctuations of interaction energy, while the imaginary part represents the spectrum of energy flow from one oscillator to another.

### 3. Spectra of oscillators

In the absence of noise, the stationary state of the system of Eq. (1) is zero. The dynamics of the system when it tends to the stationary state is determined only by the eigenvalues and eigenstates of matrix $\hat{M}$, which have the form:

$$\lambda_{1,2} = -\frac{\gamma_1 + \gamma_2}{2} \pm \frac{1}{2}\sqrt{(\gamma_1 - \gamma_2)^2 - 4\Omega^2} \qquad (7)$$

and

$$\vec{e}_{1,2} = \left\{ \begin{array}{c} \frac{i}{2\Omega}\left(\gamma_2 - \gamma_1 \pm \sqrt{(\gamma_1 - \gamma_2)^2 - 4\Omega^2}\right) \\ 1 \end{array} \right\}. \qquad (8)$$

In this system, there is an EP at which the eigenvalues are equal to each other and the eigenstates coincide. The EP arises when $\Omega = \Omega_{EP} = |\gamma_2 - \gamma_1|/2$, and separates the weak ($\Omega < \Omega_{EP}$) and strong ($\Omega > \Omega_{EP}$) coupling regimes [3,4].

It is assumed that the transition to the strong coupling regime can be detected based on the splitting in the system spectrum [3,4]. However, splitting in the fluctuation spectra of the first and second oscillators occurs at coupling strengths of $\Omega_1^{split}$ and $\Omega_2^{split}$, which differ from the coupling strength at the EP ($\Omega_{EP}$) (Fig. 1). $\Omega_{1,2}^{split}$ can be calculated from Eq. (6) and depend on the ratio of the reservoir temperatures ($T_2/T_1$) (Fig. 2). It should be noted that the splitting in the oscillator spectra can take place even when $\Omega < \Omega_{EP}$, i.e., in the weak coupling regime (see the dashed red line in Fig. 2).

Thus, the splitting point in the system spectrum depends on the reservoir temperatures, and generally does not coincide with the EP. We note that at this stage, it is difficult to provide a direct analogy with the standard theory of second-order phase transition, due to absence of an order parameter and the divergence of its fluctuations near the transition point. As we will see below, this analogy can be established if we consider the spectrum of the energy flow.

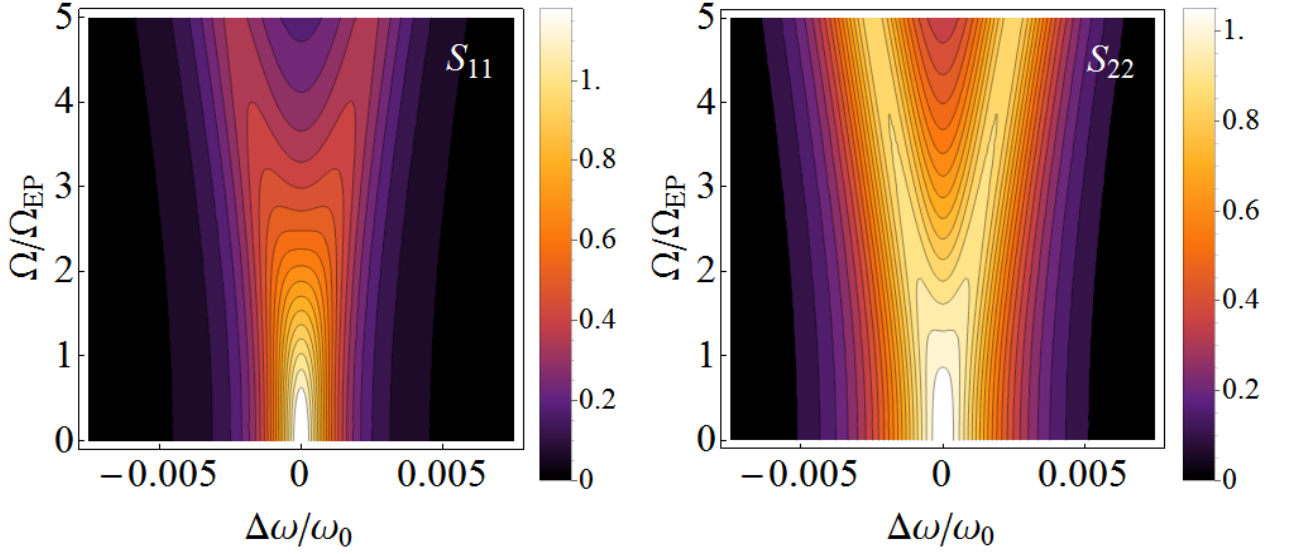

Figure 1. Spectra of the first and second oscillators for different values of the coupling strength $\Omega$. $\gamma_1 = 10^{-3}\omega_0$, $\gamma_2 = 2\times 10^{-3}\omega_0$, $T_1 = T_2$. Here, $\Omega_1^{split} \approx 2.58\,\Omega_{EP}$ and $\Omega_2^{split} \approx 1.12\,\Omega_{EP}$.

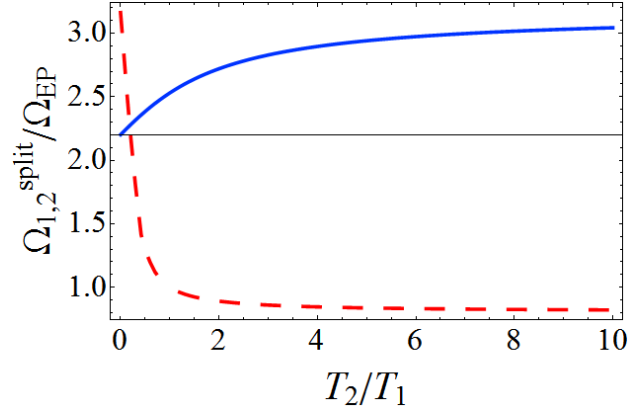

Figure 2. Dependence of $\Omega_1^{split}$ (solid blue line) and $\Omega_2^{split}$ (dashed red line) on the ratio of reservoir temperatures ($T_2/T_1$).

## 4. Non-Hermitian phase transition in the spectrum of energy flow between the reservoirs

### 4.1. Energy flow between the reservoirs

As mentioned above, there is an energy flow through the system, i.e., the system is not in thermodynamic equilibrium. The spectrum of this energy flow is determined by the imaginary parts of the non-diagonal elements of the matrix $\hat{S}(\omega)$:

$$\langle J(\omega) \rangle = \mathrm{Im}\, S_{12} = -\mathrm{Im}\, S_{21} = \frac{1}{\pi\,\Phi(\omega)}\Omega\gamma_1\gamma_2(T_2 - T_1), \qquad (9)$$

where

$$\Phi(\omega) = (\gamma_1\gamma_2 + \Omega^2)^2 - 2(\Omega^2 - \Omega_{cr}^2)\omega^2 + \omega^4 \qquad (10)$$

and

$$\Omega_{cr}^2 = \left(\gamma_1^2 + \gamma_2^2\right)/2. \tag{11}$$

The quantity $\langle J(\omega) \rangle$ represents the average energy flow between the reservoirs, and is always directed from a hotter to a colder reservoir (Eq. (9)). The amplitude of $\langle J(\omega) \rangle$ depends on the difference of the reservoir temperatures. However, the form of the spectrum $\langle J(\omega) \rangle$ does not depend on the reservoir temperatures, and is determined only by the denominator $\Phi(\omega)$. When $\Omega^2 < \Omega_{cr}^2$, there is a single maximum in the spectrum $\langle J(\omega) \rangle$ at $\omega_{max} = 0$. In the opposite case, when $\Omega^2 > \Omega_{cr}^2$, there are two maxima in the spectrum $\langle J(\omega) \rangle$ at $\omega_{max} = \pm\sqrt{\Omega^2 - \Omega_{cr}^2}$. At $\Omega^2 = \Omega_{cr}^2$, splitting arises in the spectrum of the energy flow.

Note that $\Omega_{cr} \neq \Omega_{EP}$; however, in the same way as $\Omega_{EP}$, $\Omega_{cr}$ does not depend on the reservoir temperatures. The expression for $\Phi(\omega)$ is an analog of the $\varphi^4$-potential for order parameter [40], while $\omega_{max}$ is an analog of an order parameter at the second-order phase transition. As we will show below, the fluctuations of $\omega_{max}$ diverge near $\Omega_{cr}$.

### 4.2. Critical behavior near the non-Hermitian transition

To establish the similarity between the non-Hermitian phase transition in a non-equilibrium system and the second-order phase transition, we study the fluctuation behavior of energy flow $J(\omega)$ near the critical coupling strength.

On average, the energy flow is directed from a hotter to a colder reservoir. The spectrum of the average energy flow, $\langle J(\omega) \rangle$, is determined by Eq. (9). To study the fluctuations of the energy flow, we simulate the system dynamics using Eq. (1) with noise. We calculate the stochastic time evolution of the energy flow over a finite time range $t \in \left[0, t_f\right]$, and then find the Fourier spectrum of the energy flow, $J(\omega)$. By averaging over a large number of simulations of Eq. (1), we obtain spectra for $\langle J(\omega) \rangle$ using Eq. (9) (see Fig. 3).

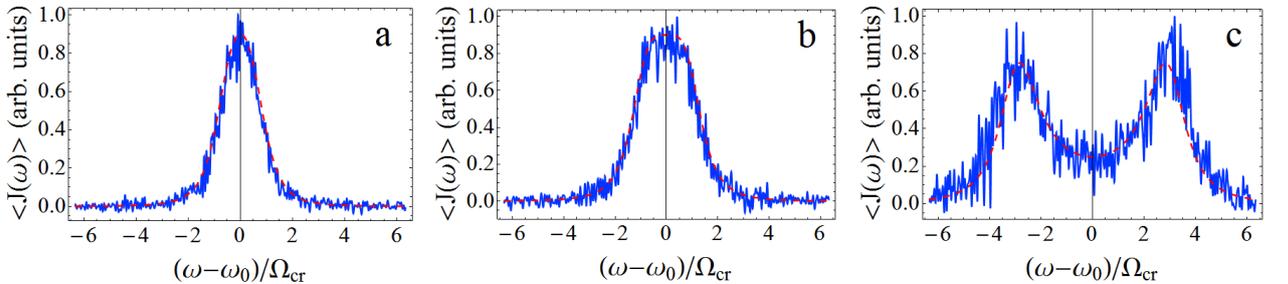

Figure 3. Spectra of the energy flow $\langle J(\omega) \rangle$, calculated by averaging over $3\times10^5$ realizations of simulations of Eq. (1) (solid blue line) and calculated using Eq. (9) (dashed red line): (a)

$\Omega = 0.67\,\Omega_{cr}$; (b) $\Omega = \Omega_{cr}$; (c) $\Omega = 3\,\Omega_{cr}$.

In a given realization, the energy flow can fluctuate and the direction of flow may change. As a result, the spectrum of energy flow calculated from a single realization differs significantly from $\langle J(\omega)\rangle$ (Fig. 4). However, the frequency distribution of the maxima in the spectrum $J(\omega)$ resembles the form of $1/\Phi(\omega)$ (cf. the solid blue and dashed red lines in Fig. 4). That is, $\Phi(\omega)$ plays the role of a potential for the distribution of maxima in the spectrum $J(\omega)$. Accordingly, the frequency of the maximum in the spectrum $J(\omega)$ can be considered as the order parameter of the non-Hermitian phase transition in a system far from equilibrium.

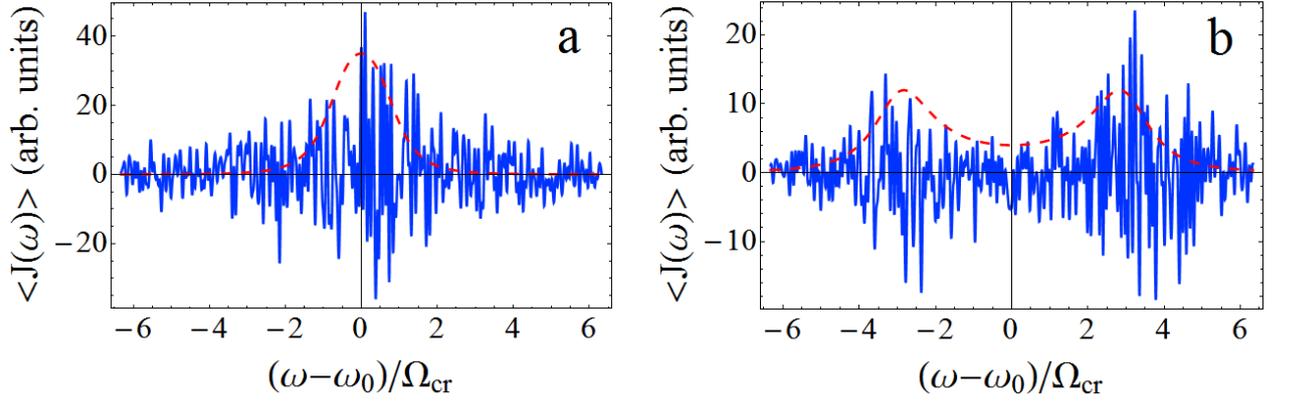

Figure 4. The spectrum of energy flow $J(\omega)$ calculated from a single simulation of Eq. (1) (solid blue line) and the $\varphi^4$-potential $\Phi(\omega)$ (dashed red line) for: (a) $\Omega = 0.67\,\Omega_{cr}$ and (b) $\Omega = 3\,\Omega_{cr}$.

To describe the fluctuations of the order parameter, we find the frequency of the maximum in the spectrum $J(\omega)$, i.e., $\omega_{max}$, for each of the simulations. We then average $\omega_{max}$ and $\omega_{max}^2$ over a large number of simulations, and calculate the dispersion $D(\omega_{max}) = \langle \omega_{max}^2 \rangle - \langle \omega_{max} \rangle^2$ for different values of the coupling strength $\Omega$ and the reservoir temperatures $T_{1,2}$ (see Fig. 5a). We can see that the behavior of $D(\omega_{max})$ changes qualitatively at the point of phase transition ($\Omega^2 = \Omega_{cr}^2$). Near the transition point, the dispersion behaves as $D(\omega_{max}) \Box |\Omega - \Omega_{cr}|^\alpha$. In accordance with the terminology used in the theory of second-order phase transitions [40], we refer to $\alpha$ as the critical exponent. It can be shown that the dispersion $D(\omega_{max})$ and the critical exponent $\alpha$ depend on the ratio of the reservoir temperatures ($T_2/T_1$) (see Fig. 5b), but do not depend on the absolute values of these temperatures (see Supplementary Materials).

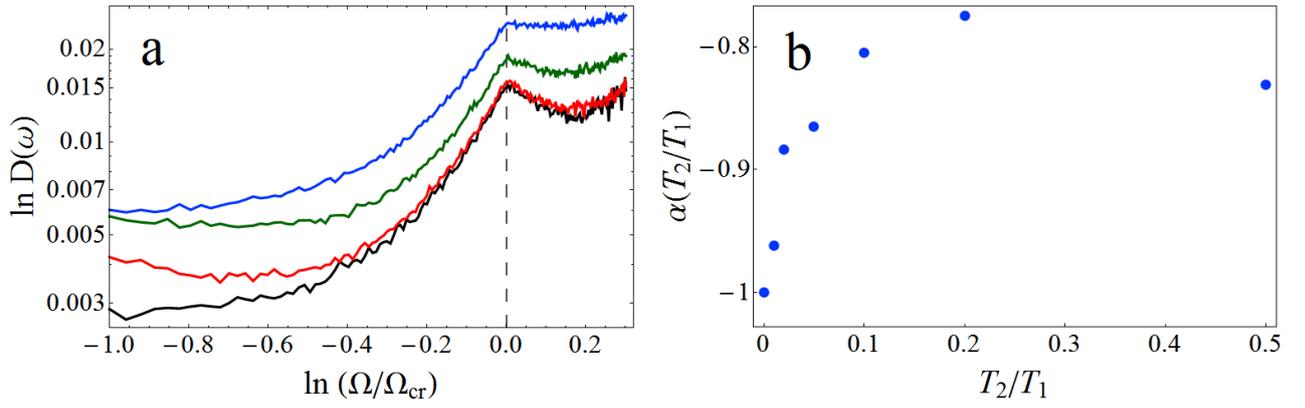

Figure 5. (a) Dependence of the dispersion $D(\omega_{max}) = \langle \omega_{max}^2 \rangle - \langle \omega_{max} \rangle^2$ on the coupling strength for different ratios of the reservoir temperatures $T_{1,2}$: $T_2/T_1 = 0$ (black line), $T_2/T_1 = 0.01$ (red line), $T_2/T_1 = 0.1$ (green line), $T_2/T_1 = 0.5$ (blue line); (b) dependence of the critical exponent on the ratio of the reservoir temperatures ($T_2/T_1$).

Note that within a small neighborhood of the transition point, the power law is violated due to the finiteness of the system, a result that also agrees with the theory of second-order phase transitions [40].

### Conclusion

We demonstrate that in non-Hermitian systems far from equilibrium, a new type of non-Hermitian phase transition takes place. We consider a system of two coupled harmonic oscillators interacting with reservoirs at different temperatures. The difference of the reservoir temperatures moves the system away from thermal equilibrium, resulting in an energy flow through the system. We show that the spectrum of this energy flow is determined by an analog of the $\varphi^4$-potential, and exhibits frequency splitting at a CP above which there are two minima in the potential. This spectral splitting at the CP can be associated with a non-equilibrium phase transition. The frequency of the maximum in the spectrum plays the role of an order parameter for this phase transition. We demonstrate that near the CP, fluctuations of the order parameter diverge according to a power law, and we calculate the critical exponent. We show that this depends only on the ratio of the reservoir temperatures, which indicates the non-equilibrium nature of the phase transition.

The non-Hermitian phase transition described here can take place in non-Hermitian systems without an EP. This opens the way for the study of non-Hermitian phase transitions in a new class of systems.

### References

[1] C. M. Bender and S. Boettcher, Phys. Rev. Lett. **80**, 5243 (1998).
[2] N. Moiseyev, *Non-Hermitian Quantum Mechanics*, (Cambridge University Press, Cambridge, UK, 2011).
[3] M.-A. Miri and A. Alù, Science **363**, eaar7709 (2019).
[4] S. Özdemir, S. Rotter, F. Nori, and L. Yang, Nat. Mater. **18**, 783 (2019).

[5]   A. A. Zyablovsky, A. P. Vinogradov, A. A. Pukhov, A. V. Dorofeenko, and A. A. Lisyansky, Phys. Usp. **57**, 1063 (2014).

[6]   R. El-Ganainy, K. G. Makris, M. Khajavikhan, Z. H. Musslimani, S. Rotter, and D. N. Christodoulides, Nat. Phys. **14**, 11 (2018).

[7]   S. Longhi, Europhys. Lett. **120**, 64001 (2018).

[8]   J. B. Khurgin, Optica **7**, 1015 (2020).

[9]   T. Gao, E. Estrecho, K. Y. Bliokh, T. C. H. Liew, M. D. Fraser, S. Brodbeck, M. Kamp, C. Schneider, S. Höfling, Y. Yamamoto, F. Nori, Y. S. Kivshar, A. G. Truscott, R. G. Dall, and E. A. Ostrovskaya, Nature **526**, 554 (2015).

[10]  D. Zhang, X. Q. Luo, Y. P. Wang, T. F. Li, and J. Q. You, Nat. Commun. **8**, 1368 (2017).

[11]  X. Zhang, K. Ding, X. Zhou, J. Xu, and D. Jin, Phys. Rev. Lett. **123**, 237202 (2019).

[12]  H. Xu, D. Mason, L. Jiang, and J. G. E. Harris, Nature **537**, 80 (2016).

[13]  J. Zhang, B. Peng, Ş. K. Özdemir, K. Pichler, D. O. Krimer, G. Zhao, F. Nori, Y.-X. Liu, S. Rotter, and L. Yang, Nature Photon. **12**, 479 (2018).

[14]  A. A. Zyablovsky, I. V. Doronin, E. S. Andrianov, A. A. Pukhov, Y. E. Lozovik, A. P. Vinogradov, and A. A. Lisyansky, Laser Photonics Rev. **15**, 2000450 (2021).

[15]  Y.-H. Lai, Y.-K. Lu, M.-G. Suh, Z. Yuan, and K. Vahala, Nature **576**, 65 (2019).

[16]  H. Hodaei, A. U. Hassan, S. Wittek, H. Garcia-Gracia, R. El-Ganainy, D. N. Christodoulides, and M. Khajavikhan, Nature **548**, 187 (2017).

[17]  W. Chen, S. K. Ozdemir, G. Zhao, J. Wiersig, and L. Yang, Nature **548**, 192 (2017).

[18]  J. Wiersig, Phys. Rev. Lett. **112**, 203901 (2014).

[19]  Z. P. Liu, J. Zhang, Ş. K. Özdemir, B. Peng, H. Jing, X. Y. Lü, C.-W. Li, L. Yang, F. Nori, and Y. X. Liu, Phys. Rev. Lett. **117**, 110802 (2016).

[20]  H. Hodaei, M.-A. Miri, M. Heinrich, D. N. Christodoulides, and M. Khajavikan, Science **346**, 975 (2014).

[21]  L. Feng, Z. J. Wong, R.-M. Ma, Y. Wang, and X. Zhang, Science **346**, 972 (2014).

[22]  B. Peng, Ş. K. Özdemir, M. Liertzer, W. Chen, J. Kramer, H. Yılmaz, J. Wiersig, S. Rotter, and L. Yang, Proc. Natl. Acad. Sci. **113**, 6845 (2016).

[23]  M. Liertzer, L. Ge, A. Cerjan, A. D. Stone, H. E. Türeci, and S. Rotter, Phys. Rev. Lett. **108**, 173901 (2012).

[24]  B. Peng, Ş. K.Özdemir, S. Rotter, H. Yilmaz, M. Liertzer, F. Monifi, C. M. Bender, F. Nori, and L. Yang, Science **346**, 328 (2014).

[25]  M. Brandstetter, M. Liertzer, C. Deutsch, P. Klang, J. Schöberl, H. E. Türeci, G. Strasser, K. Unterrainer, and S. Rotter, Nat. Commun. **5**, 4034 (2014).

[26]  I. V. Doronin, A. A. Zyablovsky, E. S. Andrianov, A. A. Pukhov, and A. P. Vinogradov, Phys. Rev. A **100**, 021801(R) (2019).

[27]  I. V. Doronin, A. A. Zyablovsky, and E. S. Andrianov, Opt. Express **29**, 5624 (2021).

[28]  S. B. Lee, J. Yang, S. Moon, S. Y. Lee, J.-B. Shim, S. W. Kim, J.-H. Lee, and K. An, Phys. Rev. Lett. **103**, 134101 (2009).

[29]  G. Q. Zhang and J. Q. You, Phys. Rev. B **99**, 054404 (2019).

[30]  A. Roy, S. Jahani, C. Langrock, M. Fejer, and A. Marandi, Nat. Commun. **12**, 835 (2021).

[31]  G. Khitrova, H. M. Gibbs, M. Kira, S. W. Koch, and A. Scherer, Nat. Phys. **2**, 81 (2006).

[32]  S. Savasta, R. Saija, A. Ridolfo, O. D. Stefano, P. Denti, and F. Borghese, ACS Nano **4**, 6369 (2010).

[33]  A. I. Vakevainen, R. J. Moerland, H. T. Rekola, A.-P. Eskelinen, J.-P. Martikainen, D.-H. Kim, and P. Torma, Nano Lett. **14**, 1721 (2014).

[34]  N. A. Mortensen, P. A. D. Gonçalves, M. Khajavikhan, D. N. Christodoulides, C. Tserkezis, and C. Wolff, Optica **5**, 1342 (2018).

[35]  H. Carmichael, *An open systems approach to quantum optics* (Springer-Verlag, Berlin, 1991).

[36]  C. W. Gardiner and P. Zoller, *Quantum noise*, (Springer Verlag, Berlin, 1991).

[37] H. Haken, *Laser light dynamics*, (North-Holland Physics Publishing Oxford, 1985).
[38] M. O. Scully and M. S. Zubairy, *Quantum optics*, (Cambridge University Press, Cambridge, 1997).
[39] L. Mandel and E. Wolf, *Optical coherence and quantum optics*, (Cambridge University Press, 1995).
[40] L. D. Landau and E. M. Lifshitz, *Statistical Physics*, (Butterworth-Heinemann, 1980).

# Supplementary Materials to "A new type of non-Hermitian phase transition in open systems far from thermal equilibrium"

T. T. Sergeev[1,2,3], A. A. Zyablovsky[1,2,3,4], E. S. Andrianov[1,2,3], A. A. Pukhov,[2,3] Yu. E. Lozovik[1,2,5], A. P. Vinogradov[1,2,3]

[1]Dukhov Research Institute of Automatics (VNIIA), 127055, 22 Sushchevskaya, Moscow, Russia

[2]Moscow Institute of Physics and Technology, 141700, 9 Institutskiy per., Moscow, Russia

[3]Institute for Theoretical and Applied Electromagnetics, 125412, 13 Izhorskaya, Moscow, Russia

[4]Kotelnikov Institute of Radioengineering and Electronics RAS, 11-7 Mokhovaya, Moscow 125009, Russia

[5]Institute of Spectroscopy Russian Academy of Sciences, 108840, 5 Fizicheskaya, Troitsk, Moscow, Russia

## 1. Derivation of the expression for the system spectrum

Using the Wiener-Khinchin theorem [1,2], we calculate the spectrum of a system

$$\hat{S}(\omega) = \frac{1}{2\pi} \int_{-\infty}^{+\infty} d\tau e^{-i\omega\tau} \left\langle \vec{a}^*(\tau)\vec{a}^T \right\rangle_{st}. \quad (S1)$$

We first calculate the correlator $\left\langle \vec{a}^*(\tau)\vec{a}^T \right\rangle_{st}$ in Eq. (S1). To do this, we formally integrate the matrix equation (1) in the main text:

$$\vec{a}(t) = \exp(\hat{M}t)\vec{a}(0) + \int_0^t d\tau \exp(\hat{M}(t-\tau))\vec{\xi}(\tau). \quad (S2)$$

Then, using the quantum regression theorem [3], we obtain:

$$\left\langle \vec{a}^*(t+\tau)\vec{a}^T(t) \right\rangle = \exp(\hat{M}^*(t+\tau))\vec{a}^*(0)\vec{a}^T(0)\exp(\hat{M}^T t) + \\ + \int_t^{t+\tau} d\tau' \int_0^t d\tau'' \exp(\hat{M}^*(t+\tau-\tau'))\left\langle \vec{\xi}^*(\tau')\vec{\xi}^T(\tau'') \right\rangle \exp(\hat{M}^T(t-\tau'')). \quad (S3)$$

Taking into account the expressions for the noise correlators (see Eq. (2) in the main text) and considering that $\vec{a}^*(0)\vec{a}^T(0) = 0$, we derive:

$$\left\langle \vec{a}^*(t+\tau)\vec{a}^T(t) \right\rangle = 2\int_0^{t+\tau} d\tau' \int_0^t d\tau'' \exp(\hat{M}^*(t+\tau-\tau'))\hat{D}\delta(\tau'-\tau'')\exp(\hat{M}^T(t-\tau'')) = \\ = 2\exp(\hat{M}^*\tau)\exp(\hat{M}^*t)(\int_0^t d\tau'' \exp(-\hat{M}^*\tau'')\hat{D}\exp(-\hat{M}^T\tau''))\exp(\hat{M}^T t) \quad (S4)$$

When $\tau \to +0$, Eq. (S4) takes the form:

$$\langle \vec{a}^*(t)\vec{a}^T(t)\rangle = 2\exp(\hat{M}^*t)(\int_0^t d\tau'' \exp(-\hat{M}^*\tau'')\hat{D}\exp(-\hat{M}^T\tau''))\exp(\hat{M}^Tt). \qquad (S5)$$

Differentiating both sides of Eq. (S5), we obtain:

$$\frac{d}{dt}\langle \vec{a}^*(t)\vec{a}^T(t)\rangle = \hat{M}^*\langle \vec{a}^*(t)\vec{a}^T(t)\rangle + \langle \vec{a}^*(t)\vec{a}^T(t)\rangle \hat{M}^T + 2\hat{D}. \qquad (S6)$$

In the steady state, $\langle \vec{a}^*(t)\vec{a}^T(t)\rangle$ is determined by the following equation:

$$\hat{M}^*\langle \vec{a}^*\vec{a}^T\rangle_{st} + \langle \vec{a}^*\vec{a}^T\rangle_{st}\hat{M}^T + 2\hat{D} = 0. \qquad (S7)$$

To calculate the correlator $\langle \vec{a}^*(\tau)\vec{a}^T\rangle_{st}$, which is determined as $\langle \vec{a}^*(\tau)\vec{a}^T\rangle_{st} = \langle \vec{a}^*(t+\tau)\vec{a}^T(t)\rangle_{t\to\infty}$, we use Eq. (S4):

$$\langle \vec{a}^*(\tau)\vec{a}^T\rangle_{st} = \exp(\hat{M}^*\tau)\langle \vec{a}^*\vec{a}^T\rangle_{st}, \tau > 0. \qquad (S8)$$

The expression in Eq. (S8) holds true when $\tau > 0$. To use the Wiener-Khinchin theorem [1,2] we need to calculate the correlator at $\tau < 0$. This correlator is calculated in the same way, as follows:

$$\langle \vec{a}^*(\tau)\vec{a}^T\rangle_{st} = \langle \vec{a}^*\vec{a}^T\rangle_{st}\exp(-\hat{M}^T\tau), \tau < 0. \qquad (S9)$$

Thus, the spectrum can be expressed as:

$$\hat{S}(\omega) = \frac{1}{2\pi}\int_{-\infty}^{+\infty} d\tau e^{-i\omega\tau}\langle \vec{a}^*(\tau)\vec{a}^T\rangle_{st} = \frac{1}{2\pi}\langle \vec{a}^*\vec{a}^T\rangle_{st}\int_{-\infty}^{0} d\tau \exp((-\hat{M}^T - i\omega\hat{I})\tau) +$$

$$+\frac{1}{2\pi}\int_{-\infty}^{0} d\tau \exp((\hat{M}^* - i\omega\hat{I})\tau)\langle \vec{a}^*\vec{a}^T\rangle_{st} =$$

$$= \frac{1}{2\pi}\int_0^{+\infty} d\tau(\langle \vec{a}^*\vec{a}^T\rangle_{st}\exp((\hat{M}^T + i\omega\hat{I})\tau) + \exp((\hat{M}^* - i\omega\hat{I})\tau)\langle \vec{a}^*\vec{a}^T\rangle_{st}) = \qquad (S10)$$

$$= -\frac{1}{2\pi}(\langle \vec{a}^*\vec{a}^T\rangle_{st}(\hat{M}^T + i\omega\hat{I})^{-1} + (\hat{M}^* - i\omega\hat{I})^{-1}\langle \vec{a}^*\vec{a}^T\rangle_{st})$$

By multiplying the spectrum matrix on the left and right by the respective matrices and taking into account Eqs. (S8) and (S9), we obtain:

$$(\hat{M}^* - i\omega\hat{I})\hat{S}(\omega)(\hat{M}^T + i\omega\hat{I}) =$$

$$= -\frac{1}{2\pi}((\hat{M}^* - i\omega\hat{I})\langle \vec{a}^*\vec{a}^T\rangle_{st} + \langle \vec{a}^*\vec{a}^T\rangle_{st}(\hat{M}^T + i\omega\hat{I})) = \qquad (S11)$$

$$= -\frac{1}{2\pi}(\hat{M}^*\langle \vec{a}^*\vec{a}^T\rangle_{st} + \langle \vec{a}^*\vec{a}^T\rangle_{st}\hat{M}^T) =_{(S7)} \hat{D}/\pi$$

Finally, we obtain an expression for the spectrum matrix:

$$\hat{S}(\omega) = \frac{1}{\pi}(\hat{M}^* - i\omega\hat{I})^{-1}\hat{D}(\hat{M}^T + i\omega\hat{I})^{-1}. \tag{S12}$$

By substituting into Eq. (S12) the expressions for the matrices $\hat{M}$ and $\hat{D}$ from the main text, we obtain:

$$\hat{S}(\omega) = \frac{1/\pi}{\Phi(\omega)}\begin{pmatrix} \gamma_1 T_1(\gamma_2^2 + \omega^2) + \gamma_2 T_2 \Omega^2 & i\Omega(\gamma_2 T_2(\gamma_1 + i\omega) - \gamma_1 T_1(\gamma_2 - i\omega)) \\ i\Omega(\gamma_1 T_1(\gamma_2 + i\omega) - \gamma_2 T_2(\gamma_1 - i\omega)) & \gamma_2 T_2(\gamma_1^2 + \omega^2) + \gamma_1 T_1 \Omega^2 \end{pmatrix}, \tag{S13}$$

where $\Phi(\omega) = (\gamma_1 \gamma_2 + \Omega^2)^2 - 2(\Omega^2 - \Omega_{cr}^2)\omega^2 + \omega^4$ and $\Omega_{cr}^2 = (\gamma_1^2 + \gamma_2^2)/2$.

The diagonal elements of $\hat{S}(\omega)$ define the spectra of the first and second oscillators, and the non-diagonal elements are complex quantities. The real parts of the non-diagonal elements are equal to each other, and define the spectrum of interaction between the oscillators, while the imaginary parts of the non-diagonal elements differ from each other in sign and define the spectra of the energy flow from the first oscillator to the second and vice versa.

Note that $\omega$ is the difference between the oscillation frequency and the frequency of a single oscillator $\omega_0$ (we consider that the frequencies of oscillators are equal to each other).

**2. Dependence of the dispersion of the order parameter on the ratio of the reservoir temperatures**

Numerical simulation of Eq. (1) shows that the spectra of the oscillators, the spectra of the interaction and the energy flow between the oscillators, and the dispersion of the frequency of the maximum in the spectrum ($\omega_{max}$) of energy flow $J(\omega)$ depend on the ratio of the reservoir temperatures rather than their absolute values. To ascertain the mechanism of this dependence, we analyze the behavior of the system with noise, as follows:

$$\frac{d}{dt}\begin{pmatrix} a_1 \\ a_2 \end{pmatrix} = \begin{pmatrix} -\gamma_1 & -i\Omega \\ -i\Omega & -\gamma_2 \end{pmatrix}\begin{pmatrix} a_1 \\ a_2 \end{pmatrix} + \begin{pmatrix} \xi_1(t) \\ \xi_2(t) \end{pmatrix} = \hat{M}\vec{a} + \vec{\xi}(t), \tag{S14}$$

where $\vec{a} = (a_1 \ a_2)^T$ is a vector of the amplitudes of the first and second oscillators, respectively; $\Omega$ is the coupling strength between the oscillators; $\gamma_{1,2}$ are the relaxation rates of the oscillators; and $\vec{\xi} = (\xi_1 \ \xi_2)^T$ is a noise term that obeys the following conditions

$$\langle \vec{\xi} \rangle = 0, \quad \langle \vec{\xi}^*(t+\tau)\vec{\xi}^T(t) \rangle = 2\hat{D}\delta(\tau), \tag{S15}$$

where $\hat{D} = \begin{pmatrix} \gamma_1 T_1 & 0 \\ 0 & \gamma_2 T_2 \end{pmatrix}$ is a diffusion matrix.

In the stationary state, the amplitudes $a_{1,2}$ are nonzero only when at least one of the reservoir temperatures is nonzero. Without loss of generality, we assume that $T_2 \neq 0$. In this case, we can make changes to the variables $\tilde{a}_{1,2} = a_{1,2}/\sqrt{T_2}$ and $\tilde{\xi}_{1,2} = \xi_{1,2}/\sqrt{T_2}$. In these new variables, the Eq. (S14) can be rewritten as:

$$\frac{d}{dt}\tilde{\vec{a}} = \hat{M}\tilde{\vec{a}} + \tilde{\vec{\xi}}(t).  \tag{S16}$$

The redefined noise terms obey the conditions

$$\left\langle \tilde{\vec{\xi}} \right\rangle = 0, \quad \left\langle \tilde{\vec{\xi}}^*(t+\tau)\tilde{\vec{\xi}}^T(t) \right\rangle = 2\tilde{\hat{D}}\delta(\tau),  \tag{S17}$$

where $\tilde{\hat{D}} = \begin{pmatrix} \gamma_1 T_1/T_2 & 0 \\ 0 & \gamma_2 \end{pmatrix}$ is a redefined diffusion matrix. It is important to note that $\tilde{\hat{D}}$, and consequently the amplitudes $\tilde{\vec{a}} = (\tilde{a}_1 \ \tilde{a}_2)^T$, depend only on the ratio of the reservoir temperatures. As a result, the spectrum matrix $\tilde{\hat{S}}(\omega)$ calculated by Eq. (S16) depends on the ratio of the reservoir temperatures (where the method of calculation is the same as for Eq. (S14)). The spectrum matrix of the initial expression in Eq. (14) is determined by $\tilde{\hat{S}}(\omega)$, as follows (see the definition of $\hat{S}(\omega)$ in Eq. (S1)):

$$\hat{S}(\omega) = T_2 \tilde{\hat{S}}(\omega).  \tag{S18}$$

Thus, the form of spectrum $\hat{S}(\omega)$ depends only on the ratio of the reservoir temperatures, while the absolute value of the temperature $T_2$ is determined only by the amplitude of the spectrum. The same conclusion holds true for a spectrum calculated based on a single simulation of Eq. (S14). As a result, the frequency of the maximum in the spectrum ($\omega_{max}$) calculated for this simulation also depends only on the ratio of the reservoir temperatures. Since the dispersion of the frequency $D(\omega_{max}) = \left\langle \omega_{max}^2 \right\rangle - \left\langle \omega_{max} \right\rangle^2$ is calculated by averaging $\omega_{max}$ and $\omega_{max}^2$ over a large number of simulations, then $D(\omega_{max})$ depends only on the ratio of the reservoir temperatures.

[1]   M. O. Scully and M. S. Zubairy, *Quantum optics*, (Cambridge University Press, Cambridge, 1997).
[2]   L. Mandel and E. Wolf, *Optical coherence and quantum optics*, (Cambridge University Press, 1995).
[3]   H. Carmichael, *An open systems approach to quantum optics* (Springer-Verlag, Berlin, 1991).